\documentclass[12pt]{article}
\usepackage{amssymb}
\usepackage{graphicx}
\usepackage[cp1251]{inputenc}
\usepackage[russian]{babel}
\usepackage{rotating}
\begin{document}
 \noindent {\footnotesize\it Astronomy Letters, 2025, Vol. 51, No. 11--12}
 \newcommand{\dif}{\textrm{d}}

 \noindent
 \begin{tabular}{llllllllllllllllllllllllllllllllllllllllllllll}
 & & & & & & & & & & & & & & & & & & & & & & & & & & & & & & & & & & & & & &\\\hline\hline
 \end{tabular}

 \vskip 0.2cm
 \bigskip
   \centerline{\bf  Analysis of the Space Velocities of Several Samples of Open Star Clusters}
   \bigskip
\centerline{V. V. Bobylev and   A. T. Bajkova  }
 \bigskip
 \centerline {\small\it Central (Pulkovo) Astronomical Observatory, Russian Academy of Sciences}
 \bigskip
{\bf Abstract}---An analysis of the kinematics of open star clusters (OSCs) using their characteristics from the new Hunt and Reffert catalog was conducted. Based on 4003 OSCs younger than 200 million years, the following values for the angular velocity of the Galaxy's rotation were found:
$\Omega_0 = 28.99\pm0.11$~km/s/kpc,
$\Omega^{'}_0 = -3.909\pm0.026$~km/s/kpc$^{2}$ and
$\Omega^{''}_0 = 0.5662\pm0.018$~km/s/kpc$^{3}$, where $V_0 = 234.8\pm3.0$~km/s for $R_0 = 8.1\pm0.1$~kpc.
It was found that periodicity in the radial velocities of OSCs is manifested in clusters younger than 600 Myr, while a wave in residual tangential velocities is observed only in the youngest ones, younger than 40 Myr. A spectral Fourier analysis of the radial velocities of three OSC samples with average ages of 18, 72, and 143 Myr was used to obtain the following values of the wavelength $\lambda$ and the velocity perturbation amplitude $f_R$:
$\lambda=2.0$~kpc and $f_R=4.3$~km/s,
$\lambda=2.2$~kpc and $f_R=8.2$~km/s,
$\lambda=2.1$~kpc and $f_R=9.6$~km/s, respectively. A systematic change in the positions of the maxima and minima of the waves in the radial velocities of OSCs was found depending on the age of the sample. From the analysis of these shifts, the value of the absolute value of the difference $|\Delta\Omega|$ between the angular velocity of rotation of the spiral pattern $\Omega_p$ and the rotation velocity of the Galaxy was found, $|\Delta\Omega|=2.0\pm0.5_{stat}\pm2.3_{syst}$~km/s/kpc. Based on this, an estimate of two possible values of the corotation radius was obtained: $8.6\pm0.2$~kpc and $7.6\pm0.2$~kpc, which indicates that the Sun is very close to the corotation.

\bigskip
 \section{INTRODUCTION}
Star groupings such as OB associations, open star clusters (OSCs), and moving clusters are of great interest for studying the structure and kinematics of the Galaxy on various scales. This is because kinematic analysis requires the most accurate possible estimates of the distances to objects and their velocities. Typically, OSCs contain approximately $10^3$ member stars with known photometry, allowing average values of cluster kinematic and photometric characteristics to be obtained with higher accuracy than for single stars. OSCs and OB associations are distributed over a wide region of the Galaxy and are therefore important for studying its large-scale structure. Age estimates obtained either from isochrone fitting or dynamical methods enable OSCs to play an important role in developing the theory of stellar evolution, testing star formation models, etc.

A huge number of scientific publications are devoted to stellar groups. For example, OSCs and OB associations, along with hydrogen clouds, OB field stars, masers in massive regions of active star formation, and Cepheids, are used to estimate the parameters of the Galactic rotation curve (Zabolotskikh et al. 2002; Piskunov et al. 2006; Bobylev et al. 2007; Bobylev and Bajkova 2016, 2023; Loktin and Popova 2019), to refine the characteristics of the galactic spiral density wave (Amaral and Lepin 1997; Popova and Loktin 2005; Loktin and Popova 2007; Bobylev et al. 2008; Lepin et al. 2008; Junqueira et al. 2015; Camargo et al. 2015; Bobylev and Bajkova 2019; Cantat-Gaudin et al. 2020), to solve other stellar-astronomical problems (e.g., Tarrick et al. 2021; Barnes et al. 2026).

Recently, with the advent of massive star catalogs, the number of discovered and studied OSCs has increased significantly (Kharchenko et al. 2005, 2007, 2013; Scholz et al. 2015; Cantat-Gaudin et al. 2018; Dias et al. 2001, 2006, 2021; Hao et al. 2021). While recent general catalogs of open clusters contained 600--1700 members (Kharchenko et al. 2007; Cantat-Gaudin et al. 2018; Liu and Pang 2019; Dias et al. 2006; 2021) with precise positions and proper motions, modern catalogs now number 4000--7000 clusters (Hao et al. 2021; Yoshi and Malhotra 2023; Hunt and Reffert 2023, 2024). However, open clusters with measured radial velocities account for only 50--60\% of their total number.

The accuracy of determining the average distances to open-clusters, their proper motions, and radial velocities is improving. Currently, the primary source of mass kinematic data on stars, such as trigonometric parallaxes, proper motions, and radial velocities, are catalogs created by the Gaia spacecraft project (Gaia Collaboration 2016).

In the current version of the catalog, Gaia DR3 (Gaia Collaboration 2023), trigonometric parallaxes for approximately 500 million stars are measured with errors of less than 0.2 milliarcseconds (mas). For stars with stellar magnitudes $G< 5^m$, random errors in proper motion measurements lie in the range of 0.02--0.04 milliarcseconds per year (mas/yr), with these errors increasing significantly for fainter stars. The proper motions of about half the stars in the catalog are measured with a relative error of less than 10\%.

The main distinguishing feature of the Gaia DR3 catalog from previous versions is that it significantly increases the number of stars with measured radial velocities. However, the radial velocity measurement errors, $\sigma_{V_r}$, depend strongly on stellar magnitude. For example, Fig. 4 of Babusiaux et al. (2023) shows that $\sigma_{V_r}$ do not exceed 10~km/s for stars with $G<8^m$, while for stars in the range $10^m<G<15^m$, $\sigma_{V_r}$ values cfn exceed 40~km/s. This applies to the radial velocity measurements of individual stars. Obviously, average values calculated over a large number of members of an OSC can be significantly smaller.

The Gaia project data have already made a significant contribution to the study of the Galaxy. Of great interest is the review by Perryman (2025), which describes in detail the Gaia contribution made by analyzing Solar System objects, variable stars, virtually all known structural elements of the Milky Way, and the Local Group of galaxies. We note some interesting results obtained from the Gaia data. For example, this includes the discovery of the so-called ``Gaia phase space spiral'', which was first observed in the Gaia DR2 data (Gaia Collaboration 2018) by Antoja et al. (2018). This discovery reveals the presence of periodic perturbations in the vertical velocities of stars in the galactic disk, consistent with large-scale radial perturbations of their coordinates and velocities. This field, sometimes referred to as galactic seismology, is currently being actively studied (e.g., Antoja et al. 2023; Chiba et al. 2025; Tepper-Garcia et al. 2025; Hamilton et al. 2026; Yamsiri et al. 2026).

Using proper motions from the Gaia\,DR2 catalog, a large sample of classical Cepheids, in combination with the high-precision distances obtained in Skowron et al. (2019a) based on the period-luminosity relation, Mroz et al. (2019) constructed a Galactic rotation curve over a very wide distance range of $4<R<20$~kpc. High-precision trigonometric parallaxes, proper motions, and radial velocities of stars close to the Sun (closer than $\sim$500 pc) taken from the Gaia catalog were used to create new extensive lists of members of stellar groups, open-clusters, and OB associations, such as $\beta$~Pictoris, TW~Hya, $\rho$~Oph, Lup, $\eta$~Cha, Cha\,I, Cha\,II, etc., and to estimate their various structural and dynamic characteristics using these data, in particular, new estimates of their dynamical age (e.g., Luhman 2023; Ratzenb\"ock et al. 2023; Lee et al. 2024; Armstrong et al. 2025). Finally, Gaia parallaxes and photometry were used to create new interstellar extinction maps in a region approximately 2 kpc from the Sun (e.g., Green et al. 2019; Gontcharov et al. 2025).

Bobylev and Bajkova (2023) showed that open-clusters younger than 50 million years are strongly influenced by the galactic spiral density wave, both radially and tangentially. Data on these clusters were taken from the Hunt and Reffert (2023) catalog. In this paper, we aim to redefine the galactic rotation parameters and, using radial and residual tangential velocities, trace the age of open-clusters that are influenced by the galactic spiral density wave. This is especially true now that a new, improved version of the Hunt and Reffert (2024) catalog has been released.

\begin{figure}[t]{ \begin{center}
\includegraphics[width=0.95\textwidth]{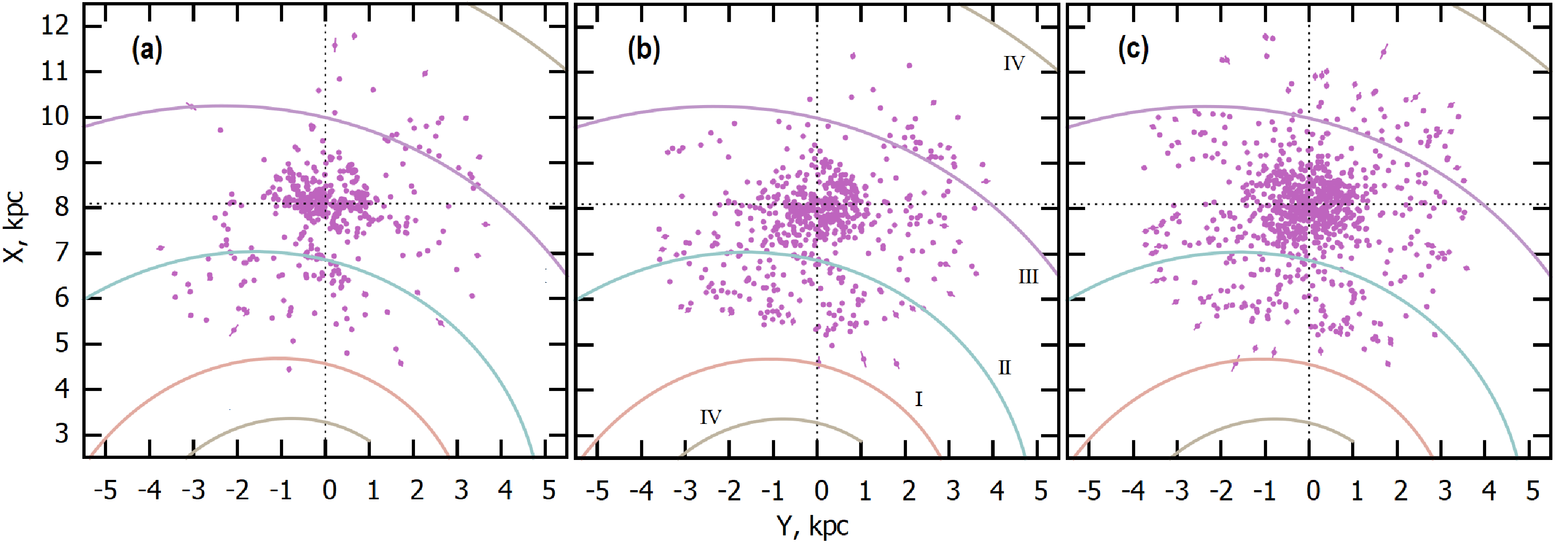}
\caption{Positions of OSCs younger than 40 million years projected onto the galactic $XY$ plane (a), with ages in the range of 40--100 million years (b) and with ages in the range of 100--200 million years (c); a four-arm spiral pattern with a pitch angle $i=-13^\circ$ is shown, constructed according to the work of Bobylev and Bajkova (2014).}
\label{fig-spiral}\end{center}}
\end{figure}

\section{DATA}
The papers by Hunt and Reffert (2021, 2023, 2024) are devoted to the creation of a catalog of open-clusters based on data from the Gaia (2016) project. In Hunt and Reffert (2021), a method for identifying cluster members using the HDBSCAN cluster analysis algorithm is described in detail, where stellar data were taken from the Gaia\,DR2 version (Gaia Collab. 2018). In Hunt and Reffert (2023, 2024), kinematic and photometric stellar data were taken from the Gaia\,DR3 catalog (Gaia Collab. 2023).

The current version of the Hunt and Reffert (2024) catalog includes 5,647 OSCs and 1,309 moving groups. Of these, 3,530 OSCs and 539 moving groups are of high quality, with most located at heliocentric distances less than 4 kpc. The catalog provides the average values of the OSC trigonometric parallaxes, proper motions, and radial velocities. Cluster age estimates obtained by these authors using the isochrone fitting are also provided.

It should be noted that Hunt and Reffert (2023, 2024) provide cluster distances using the Gaia parallax zero-point correction of $\Delta\pi = -0.017$~mas, as defined by Lindegren et al. (2021). In this paper, we use precisely these distances, copied from columns 600--614 of the Hunt, Reffert~(2024) catalog and the corresponding heliocentric rectangular coordinates $x,y,z$ from columns 641--690. We select objects with the marks `o' and `m', i.e., open-clusters and moving groups.

Fig.~\ref{fig-spiral} shows the distribution of selected OSC younger than 40 Myr, in the range of 40--100 Myr, and in the range of 100--200 Myr, projected onto the galactic plane $XY$. The coordinate system used is one in which the $X$ axis is directed from the galactic center to the Sun, and the $Y$ axis coincides with the direction of galactic rotation. A four-arm spiral pattern with a pitch angle $i=-13^\circ$ (Bobylev, Bajkova, 2014), constructed with the value $R_0=8.1$~kpc, is shown. The following segments of the spiral arms are numbered with Roman numerals: I~--- Scuti, II~--- Carina-Sagittarius, III~--- Perseus, and IV~--- Outer Arm.

\section{METHODS}
To estimate the parameters of galactic rotation, we use a system of conditional equations obtained
by expanding the angular velocity of the Galaxy's rotation $\Omega$ in a series up to terms of the second order of smallness $r/R_0:$
\begin{equation}
\begin{array}{lll}
V_r=-U_\odot\cos b\cos l-V_\odot\cos b\sin l -W_\odot\sin b\\
+R_0(R-R_0)\sin l\cos b\Omega^\prime_0+0.5R_0(R-R_0)^2\sin l\cos b\Omega^{\prime\prime}_0,
\label{EQ-1}
\end{array}
\end{equation}
\begin{equation}
\begin{array}{lll}
V_l= U_\odot\sin l-V_\odot\cos l-r\Omega_0\cos b\\
+(R-R_0)(R_0\cos l-r\cos b)\Omega^\prime_0+0.5(R-R_0)^2(R_0\cos l-r\cos b)\Omega^{\prime\prime}_0,
\label{EQ-2}
\end{array}
\end{equation}
\begin{equation}
\begin{array}{lll}
V_b=U_\odot\cos l\sin b + V_\odot\sin l \sin b -W_\odot\cos b\\
-R_0(R-R_0)\sin l\sin b\Omega^\prime_0-0.5R_0(R-R_0)^2\sin l\sin b\Omega^{\prime\prime}_0,
\label{EQ-3}
\end{array}
\end{equation}
where $V_r$ is the radial velocity of the OSC, $V_l$ is the component of the OSC velocity along the galactic longitude, $V_b$ is the component of the OSC velocity along the galactic latitude, $R$ is the distance of the OSC from the Galactic rotation axis $R^2=r^2\cos^2 b-2R_0 r\cos b\cos l+R^2_0.$ The velocities $(U,V,W)_\odot$ are the average group velocity of the sample, are taken with the opposite sign and reflect the peculiar motion of the Sun. $\Omega_0$ is the angular velocity of rotation of the Galaxy at solar distance $R_0$, the parameters $\Omega^{\prime}_0$ and $\Omega^{\prime\prime}_0$ are the corresponding derivatives of the angular velocity.

The linear velocity of the Galaxy's rotation at solar distance is $V_0=|R_0\Omega_0|$. In this paper, we use the value $R_0=8.1\pm0.1$~kpc, as per the review by Bobylev and Bajkova (2021), where it was derived as a weighted average from a large number of individual estimates.

The system of conditional equations of the form~(\ref{EQ-1})--(\ref{EQ-3}) is solved by the least squares method (LSM) for six unknowns $U_\odot,$ $V_\odot,$ $W_\odot,$ $\Omega_0,$ $\Omega^{\prime}_0$ and $\Omega^{\prime\prime}_0$ with weights of the form $w_{r,l,b}=S_0/\sqrt {S_0^2+\sigma^2_{V_r, V_l, V_b}},$ where $S_0$ is the ``cosmic'' dispersion, the value of which is assumed to be equal to 10~km/s, and $\sigma_{V_r}, \sigma_{V_l}, \sigma_{V_b}$ are the errors in the corresponding observed velocities. The least squares solution is sought in several iterations using the $3\sigma$ criterion to eliminate large residuals.

Two OCL velocities are of interest: the radial velocity $V_R$, directed from the galactic center toward the Sun, and the orthogonal tangential velocity $V_{\rm circ}$, directed in the direction of the galaxy's rotation. They can be found as follows.
 \begin{equation}
 \begin{array}{lll}
  V_{\rm circ}= U\sin \theta+(V_0+V)\cos \theta, \\
           V_R=-U\cos \theta+(V_0+V)\sin \theta,
 \label{VRVT}
 \end{array}
 \end{equation}
Where the position angle $\theta$ satisfies the relation $\tan\theta=y/(R_0-x)$, $x,y,z$ are the rectangular
heliocentric coordinates of the OSC (the velocities $U,V,W$ are directed along the corresponding axes $x,y,z$). To analyze the periodicities in the tangential velocities, it is necessary to subtract the galactic rotation curve from them and form the residual velocities $\Delta V_{\rm circ}$.

To study the periodicities in the radial $V_R$ and residual tangential velocities $\Delta V_{\rm circ}$, we use spectral Fourier analysis. This method is described in detail in Bajkova and Bobylev (2012). Its main feature is that it takes into account the position angles of objects and the logarithmic nature of the spiral wave.

\begin{figure}[t]{ \begin{center}
\includegraphics[width=0.75\textwidth]{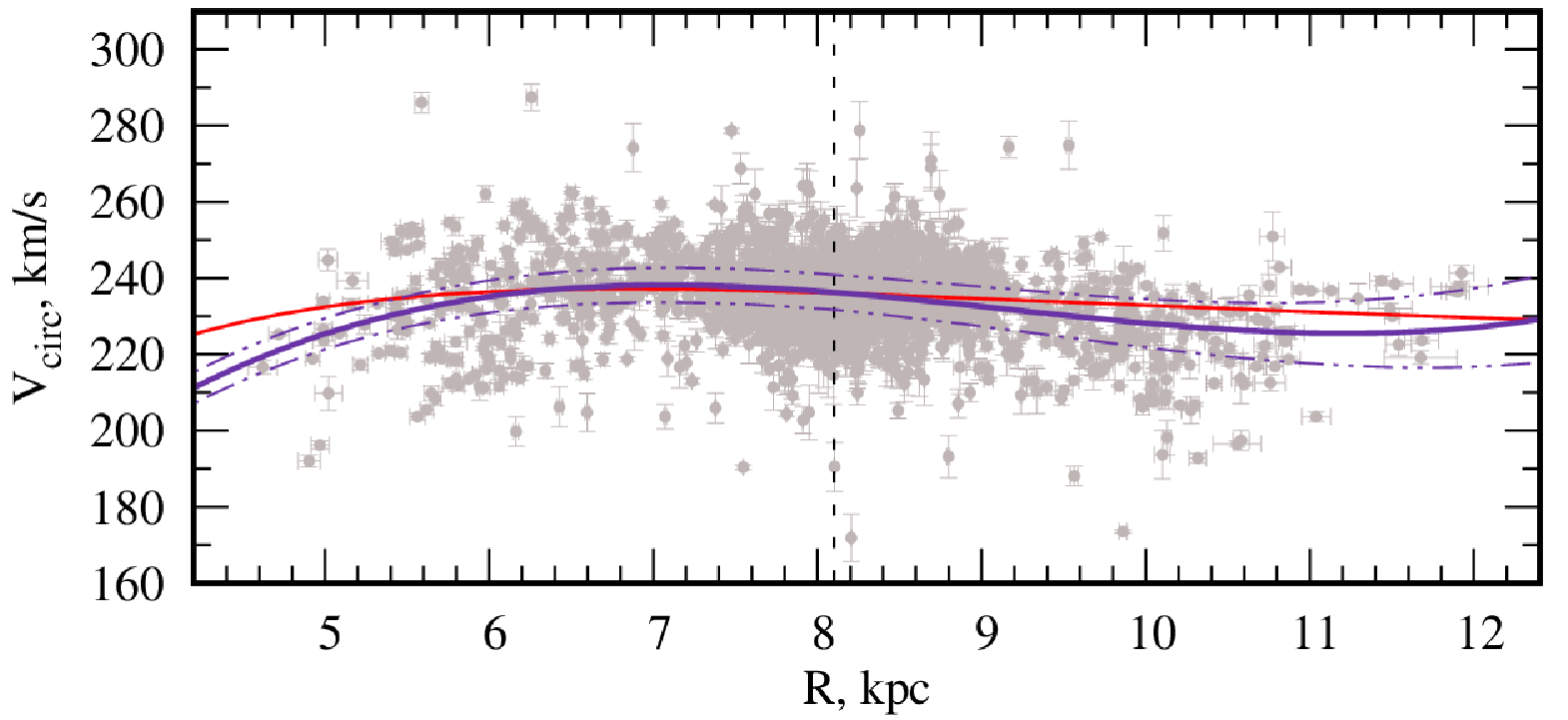}
\caption{The purple line is the Galactic rotation curve constructed in this paper based on OSCs younger than 200 Myr; the dash-dotted lines mark the boundaries of the confidence region corresponding to the 1$\sigma$ level; the red line is the curve constructed by Reid et al. (2019) based on masers with trigonometric parallaxes measured by the VLBI method, The vertical dashed line marks the position of the Sun.}
\label{fig-rot curve}\end{center}}
\end{figure}
\begin{figure}[p]{ \begin{center}
\includegraphics[width=0.5\textwidth]{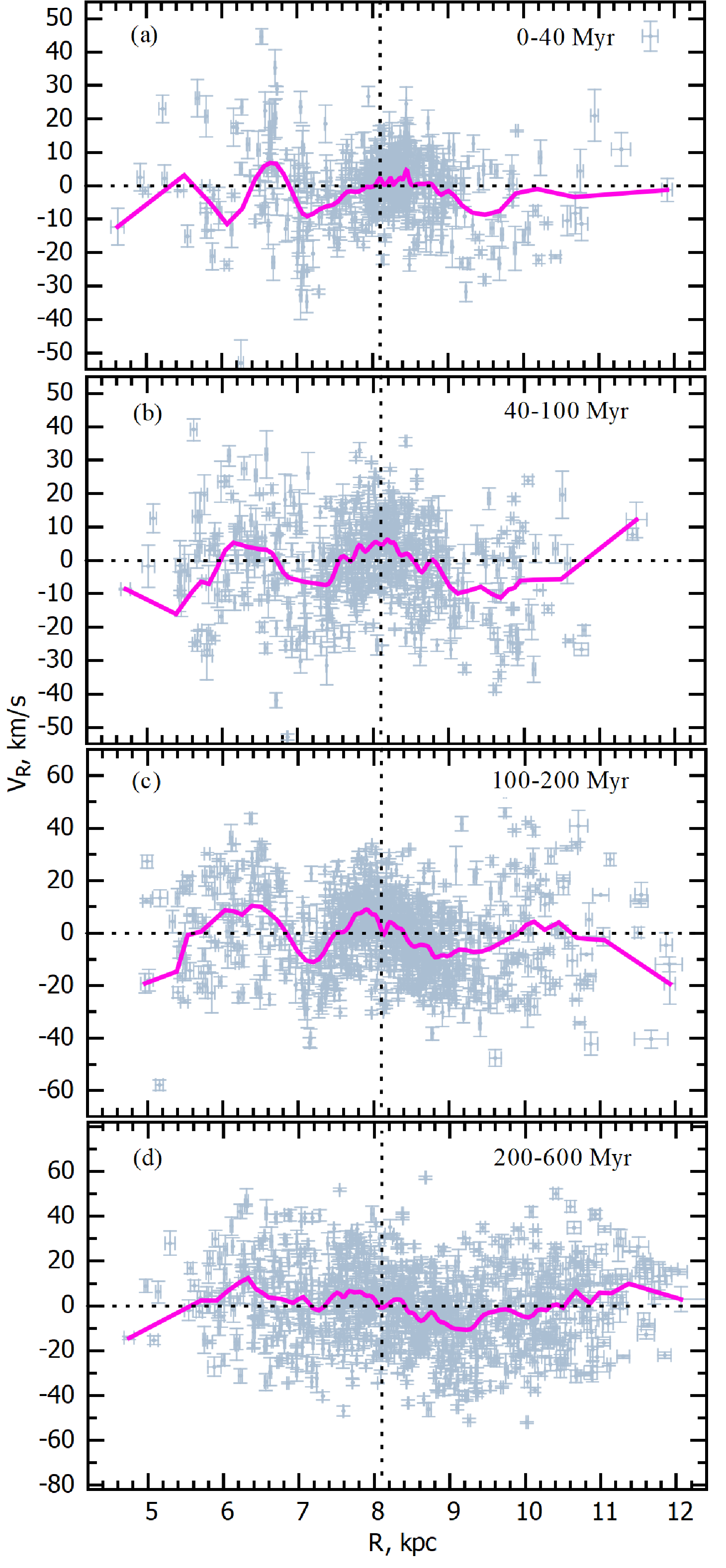}
\caption{Radial velocities $V_{R}$ of four OSC samples with average ages of 18 Myr (a), 72 Myr (b), 143 Myr (c), and 340 Myr (d) as a function of distance $R$, averaged data~--- purple thick line; The dashed line marks the position of the Sun.}
\label{fig-Rad}\end{center}}
\end{figure}
\begin{figure}[t]{ \begin{center}
\includegraphics[width=0.5\textwidth]{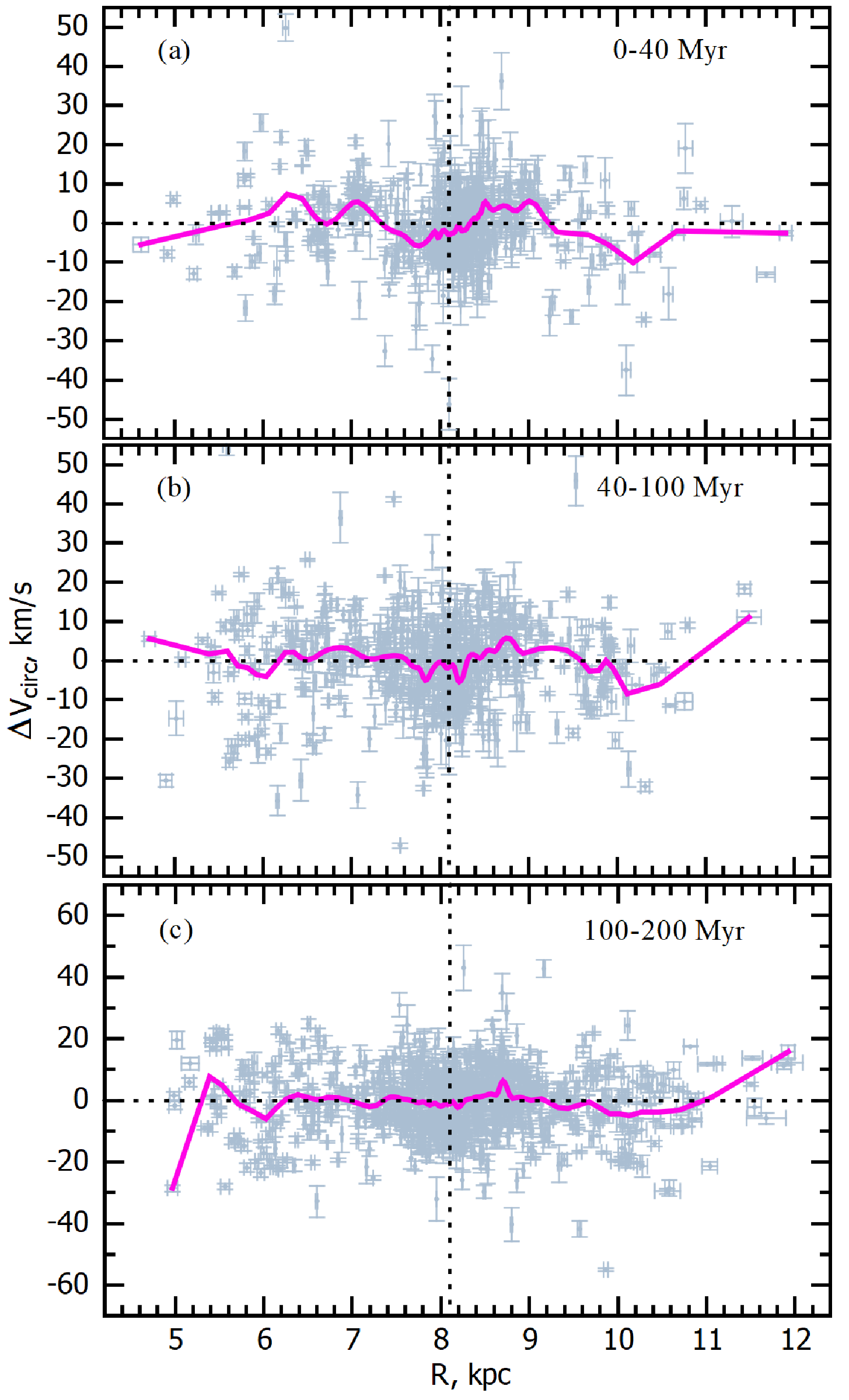}
\caption{Residual tangential velocities $\Delta V_{circ}$ of three OSC samples with average ages of 18 Myr (a), 72 Myr (b), and 143 Myr (c) as a function of the galactocentric distance $R$, averaged data~--- purple thick line; The dotted line marks the position of the Sun.}
\label{fig-Circ}\end{center}}
\end{figure}

\begin{table}[t] \caption[]{\small
Galactic rotation parameters, determined from OSC of different ages using only their proper motions
(equations~(\ref{EQ-2})--(\ref{EQ-3})), $N_\star$ --- number of clusters used, $\overline {\rm Age}$ --- average age of the sample.
}
\begin{center} \label{t:01}
\small
\begin{tabular}{|l|r|r|r|r|r|r|}\hline
Parameters & $<$40 Myr & 40--100 Myr & 100--200 Myr & 200--600 Myr \\\hline
$N_\star$ & 600 & 611 & 916 & 960\\
$\overline {\rm Age},$ Myr & 18 & 72 & 143 & 340\\
$U_\odot,$ km/s & $ 8.48\pm0.32$ & $ 8.98\pm0.40$ & $ 9.13\pm0.36$ & $ 9.83\pm0.42$\\
$V_\odot,$ km/s & $11.86\pm0.35$ & $12.52\pm0.48$ & $12.11\pm0.41$ & $13.33\pm0.50$\\
$W_\odot,$ km/s & $ 8.22\pm0.31$ & $ 7.70\pm0.40$ & $ 7.66\pm0.36$ & $ 7.89\pm0.41$\\

$\Omega_0,$ km/s/kpc & $29.44\pm0.29$ & $29.29\pm0.30$ & $29.00\pm0.28$ & $27.85\pm0.27 $\\
$\Omega^{'}_0,$ km/s/kpc$^{2}$ & $-4.016\pm0.061$ & $-4.062\pm0.064$ & $-4.002\pm0.058$ & $-3.971\pm0.062$\\
$\Omega^{''}_0,$ km/s/kpc$^{3}$ & $ 0.838\pm0.048$ & $ 0.666\pm0.058$ & $ 0.697\pm0.042$ & $ 0.740\pm0.043$\\

$\sigma_0,$ km/s & 7.7 & 9.9 & 10.8 & 12.7\\

$A,$ km/s/kpc & $ 16.26\pm0.32$ & $ 16.45\pm0.33$ & $16.21\pm0.31$ & $16.08\pm0.32$\\
$B,$ km/s/kpc & $-13.17\pm0.43$ & $ -12.84\pm0.41$ & $-12.79\pm0.42$ & $ -11.77\pm0.42$\\
  $V_0,$ km/s & $238.4\pm3.8$ & $237.2\pm3.8$ & $234.9\pm3.7$ & $225.6\pm3.5$ \\
\hline
\end{tabular}\end{center} \end{table}

\section{RESULTS}
Table~\ref{t:01} presents the parameters found by solving equations of the form~(\ref{EQ-1})--(\ref{EQ-3}) using the least squares method for four samples of OSC of different ages.

Table~\ref{t:01} presents the values of the Oort constants $A=-0.5R_0\Omega^\prime_0$ and $B=-\Omega_0-0.5R_0\Omega^\prime_0$, the signs of which are consistent with the fact that the angular velocity of rotation $\Omega$ is considered positive in this paper, then $A-B=\Omega_0.$

Table~\ref{t:200} presents the parameters found by solving equations of the form~(\ref{EQ-1})--(\ref{EQ-3}) using the least-squares method for OSCs younger than 200 Myr, which were obtained by solving conditional equations of the form~(\ref{EQ-1})--(\ref{EQ-3}) in three different ways. In the first case, variant $(V_l+V_b+V_R)_a$, all three equations~(\ref{EQ-1})--(\ref{EQ-3}) were used in the search for the least-squares solution, but only OSCs with available radial-velocity estimates, the measurement errors of which do not exceed 8~km/s, were included.
In the second case, variant $(V_l+V_b+V_R)_b$, all three equations~(\ref{EQ-1})--(\ref{EQ-3}) were also solved using the least-squares method, but the number of equations increased due to the inclusion of OSCs with proper motions, but without radial-velocity measurements.
In the third case, variant $V_l+V_b$, only two equations~(\ref{EQ-2}) and (\ref{EQ-3}) were used in the search for the least-squares solution. As can be seen from the table, in the third case, the values of the sought parameters $\Omega_0,$ $\Omega^{\prime}_0$ and $\Omega^{\prime\prime}_0$ were obtained by the least squares method with the smallest errors compared to other solution options.

\begin{table}[t] \caption[]{\small
Galactic rotation parameters found from OSC younger than 200 million years old, $N_\star$ --- number of clusters used, $N_{eq}$ --- number of equations.
}
\begin{center} \label{t:200}
\small
\begin{tabular}{|l|r|r|r|r|r|r|}\hline
Parameters & $(V_l+V_b+V_R)_a$ & $(V_l+V_b+V_R)_b$ & $V_l+V_b$ \\\hline
$N_\star$ & 2127 & 2936 & 4003 \\
$N_{eq}$ & 6316 & 7911 & 7992 \\
$\overline {\rm Age},$ million years & 88 & 84 & 79 \\ 
$U_\odot,$ km/s & $ 8.92\pm0.21$ & $ 8.56\pm0.19$ & $ 9.10\pm0.18$ \\
$V_\odot,$ km/s & $12.17\pm0.24$ & $11.76\pm0.22$ & $ 8.66\pm0.26$ \\
$W_\odot,$ km/s & $ 7.85\pm0.22$ & $ 7.96\pm0.18$ & $ 7.90\pm0.13$ \\

$\Omega_0,$ km/s/kpc & $29.17\pm0.17$ & $29.19\pm0.13$ & $28.99\pm0.11$ \\
$\Omega^{'}_0,$ km/s/kpc$^{2}$ & $-4.022\pm0.035$ & $-3.999\pm0.030$ & $-3.909\pm0.026$ \\
$\Omega^{''}_0,$ km/s/kpc$^{3}$ & $ 0.715\pm0.027$ & $ 0.760\pm0.022$ & $ 0.662\pm0.018$ \\

$\sigma_0,$ km/s & 9.8 & 9.5 & 8.3 \\

$A,$ km/s/kpc & $16.29\pm0.25$ & $16.20\pm0.23$ & $15.83\pm0.22$ \\
$B,$ km/s/kpc & $-12.88\pm0.30$ & $-12.99\pm0.27$ & $-13.15\pm0.25$ \\
$V_0.$ km/s & $ 236.3\pm3.2$ & $ 236.4\pm3.1$ & $ 234.8\pm3.0$ \\
\hline
\end{tabular}\end{center} \end{table}

Fig.~\ref{fig-rot curve} shows the Galactic rotation curve constructed from a sample of open clusters younger than 200 Myr. The curve parameters are taken from the first column of table~\ref{t:200}.

Fig.~\ref{fig-Rad} shows the radial velocities $V_{R}$ of four samples of clusters with different mean ages as functions of the Galactocentric distance $R$.

Fig. ~\ref{fig-Circ} shows the residual tangential velocities $\Delta V_{circ}$ for three samples of clusters with different average ages as functions of the Galactocentric distance $R$. Galactic rotation was taken into account using the parameters from the corresponding columns of Table~\ref{t:01}, which, incidentally, are very close to each other for these three samples of OSCs.
No periodicity was detected in the tangential velocities of OSCs in the 200--600 Myr age range; therefore, the corresponding graph is not shown to save space.

\begin{figure}[t]{ \begin{center}
\includegraphics[width=0.9\textwidth]{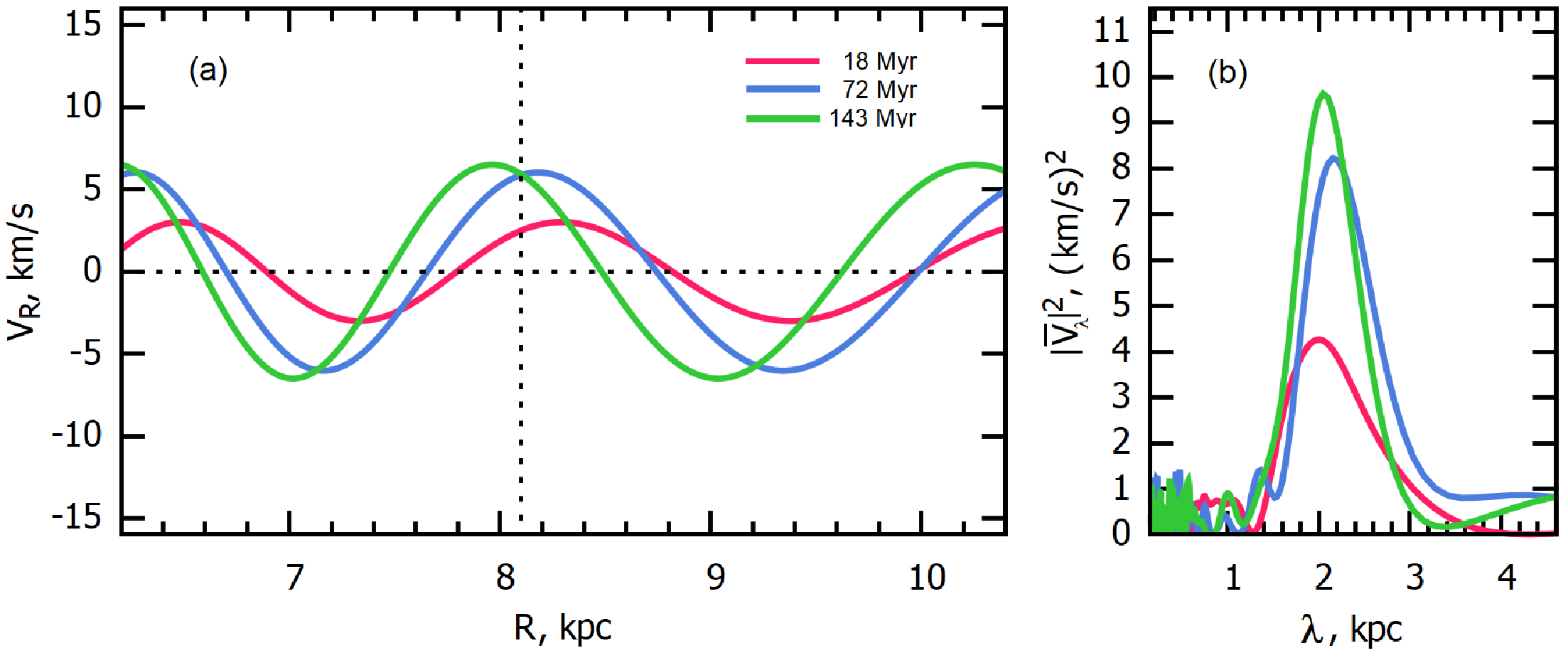}
\caption{Periodic curves found using Fourier analysis of the radial velocities $V_R$ of three samples of OSCs (a), power spectra of the radial velocities of OSCs of three ages (b).}
\label{fig-imr}\end{center}}
\end{figure}

Fig. ~\ref{fig-imr} shows the periodic curves found from the radial velocities of three samples of OSCs and their power spectra. These results were obtained through a spectral Fourier analysis of the radial velocities of the OSCs. From Fig.~\ref{fig-Rad}(d) it is evident that although the wave is visible, it has a smaller amplitude. The main difference from the others is that the wavelength here is about 7 kpc. Therefore, the spectral Fourier analysis was performed on three samples. The power spectra (Fig.~\ref{fig-imr}(b)) show good agreement in the wavelength of the disturbances $\lambda$. From this graph (Fig.~\ref{fig-imr}(b)) it is easy to find the amplitude of the disturbance velocities $f_R$, based on the relation $f_R=\sqrt {4 S_{\rm peak} }$, where $S_{\rm peak}$ is the peak value of the velocities $|{\overline V}_\lambda|^2$ in this graph.

As a result, based on the spectral Fourier analysis of radial velocities for samples of open-cluster stars with average ages of 18, 72, and 143 Myr, the following were found:
$\lambda=2.0$~kpc $f_R=4.3$~km/s and $\chi'_\odot=34.6^\circ$,
$\lambda=2.2$~kpc $f_R=8.2$~km/s and $\chi'_\odot=13.9^\circ$,
$\lambda=2.1$~kpc $f_R=9.6$~km/s and $\chi'_\odot=-23.8^\circ$, respectively. Moreover, $\chi'_\odot$ is the angle measured from the Sun's position in the direction of increasing $R$.

\section{DISCUSSION}
\subsection{Galactic Rotation Parameters}
Reid et al. (2019) performed a kinematic analysis of a sample of 147 masers with measured VLBI trigonometric parallaxes and proper motions. The uniqueness of the data lies in the fact that the maser sources taken are associated either with protostars or with very young stars. And the measurements themselves are highly accurate and absolute, tied to distant quasars, and implement an independent (from Gaia) inertial coordinate system. These authors found the following values of two most important kinematic parameters: $R_0=8.15\pm0.15$~kpc and $\Omega_\odot=30.32\pm0.27$~km/s/kpc, where $\Omega_\odot=\Omega_0+V_\odot/R,$ and the velocity value $V_\odot=12.2$~km/s was taken from the work of Sch\"onrich et al. (2010). Then we obtain $V_0=234.9\pm5.0$~km/s. The rotation curve found by these authors is presented in tabular form. It is shown by the red line in Fig.~\ref{fig-rot curve}, where we can see excellent agreement with the curve we found for a sample of young OSCs.

Bobylev and Bajkova (2022) analyzed the kinematics of OSC using the mean proper motions and distances calculated by Hao et al. (2021) using data from the Gaia~EDR3 catalog (Gaia Collaboration 2021). The least-squares solution was obtained with the lowest errors for a sample of 967 OCSs with an average age of 18 million years, using only proper motions and distances. As a result, the following values of the angular velocity parameters of the Galaxy's rotation were found:
$\Omega_0 = 28.01\pm0.15$~km/s/kpc,
$\Omega^{'}_0=-3.674\pm0.040$~km/s/kpc$^{2}$ and
$\Omega^{''}_0=0.565\pm0.023$~km/s/kpc$^{3}$, where $V_0=226.9\pm3.1$~km/s for the adopted value $R_0=8.1\pm0.1$~kpc.

Bobylev and Bajkova (2023) studied the kinematics of the Galaxy using the parameters of the OSC from the second edition of the Hunt and Reffert (2023) catalog, in which they were calculated based on the Gaia~DR3 catalog.
For 2494 OSCs younger than 50 million years, the following values were found:
$(U,V,W)_\odot=(8.43,8.70,7.88)\pm(0.24,0.27,0.22)$~km/s, and also
$\Omega_0 =29.18\pm0.12$~km/s/kpc,
$\Omega^{'}_0=-3.854\pm0.030$~km/s/kpc$^{2}$ and
$\Omega^{''}_0=0.590\pm0.012$~km/s/kpc$^{3}$, where $V_0=236.4\pm3.1$~km/s for the adopted $R_0=8.1\pm0.1$~kpc.
It should be noted that in the paper by Bobylev and Bajkova (2023), unlike the present work, the heliocentric distances of OSCs were calculated using averaged trigonometric parallax values (given in the catalog by Hunt and Reffert 2023) and no constraint was imposed on the distance $r$, although a constraint on the radius $R>4$~kpc was applied.

The paper by Feng et al. (2025) is devoted to determining the parameters of the Milky Way's rotation curve in the range 6 kpc $< R < $ 18 kpc using a sample of 903 classical Cepheids with proper motions and radial velocities from the Gaia DR3 catalog. The distances to these Cepheids were measured using the period-luminosity relation and the Wesenheit method. As a result, these authors obtained an estimate of the circular velocity of the Galaxy $V_0= 236.8\pm0.8$~km/s at a distance of $R_0$.

It is interesting to note the work of Horta et al. (2026), in which the parameters of the galactic rotation curve were found using a method that exploits the gradients of element abundances in the galactic disk. For this purpose, elemental abundance data for stars from the Apache Point Observatory Galactic Evolution Experiment (APOGEE) survey (Majewski et al. 2017) were used in combination with the kinematic properties of stars from the Gaia~DR3 catalog. The method is based on the assumption that the average values of the [Fe/H] index track the radial positions of stars in the disk, while [Mg/Fe] tracks the orbital deviations around this radius. The circular velocity of the Galaxy was found to be $V_0=235.3^{+2.8}_{-3.7}$~km/s at a distance of $R_0$, as well as
$A=16.5^{+0.1}_{-0.1}$~km/s/kpc and $B=-11.9^{+0.1}_{-0.3}$~km/s/kpc. These values are in very good agreement with the estimates of these parameters obtained in the present study for open-clusters younger than 200 Myr.

In the paper by Fedorov et al. (2025), a sample of approximately 4.5 million red giant clump stars was used to construct the Galactic rotation curve. Their spatial velocities were calculated using data from the Gaia~DR3 catalog. The rotation curve covers the range of Galactocentric distances $0<R<20$~kpc. The circular velocity of the Galaxy was found to be $V_0=229.63\pm0.30$~km/s at a distance of $R_0$. The red giant clump stars are relatively old, so the estimate obtained from them should be compared with the result of an analysis of a sample of OSCs with ages in the range of 200--600 Myrs (the last column of table \ref{t:01}), where the value obtained was $V_0=225.6\pm3.5$~km/s.

\subsection{Periodic Velocity Perturbations}
As can be seen from Figs.~\ref{fig-Rad} and \ref{fig-Circ}, there is a pronounced periodicity in the radial velocities of OSCs throughout the analyzed age range, while a wave in the residual tangential velocities is observed only in the youngest OSCs (Fig.~\ref{fig-Circ}(a)).

In the work by Bobylev and Bajkova (2023), as a result of a spectral Fourier analysis of radial and residual rotation velocities, they obtained estimates of the perturbation amplitudes $f_R=4.72\pm0.24$~km/s and $f_\theta=3.69\pm0.20$~km/s based on the spatial velocities of 1722 OSCs younger than 50~Myr, taken from the catalog of Hunt and Reffert (2023). The following estimates of the perturbation wavelength for each type of velocity were also obtained: $\lambda_R=2.4\pm0.2$~kpc and $\lambda_\theta=2.1\pm0.2$~kpc. The interpretation was that we observe the influence of the galactic spiral density wave in both the radial and tangential velocities of the OSC.

Note that the periodicity of radial velocities with an amplitude of 10--15 km/s and a wavelength of 3--4 kpc in the $R_0$ region can be seen in Fig. 2 of Fedorov et al. (2025), where these velocities are plotted based on a large sample of red giant clump stars. The radial velocities reach their maximum amplitude, approximately 25 km/s, in the inner region of the Galaxy, at $R<5 $~kpc, where the influence of the bar is strong. Furthermore, the waveform depends on the azimuthal projection. Such manifestations of radial velocities may be related to the influence of the galactic spiral density wave. The only difficulty is the absence of clearly defined periodic perturbations in the residual tangential velocities of red giant clump stars, as can be seen in Fig. 3 of Fedorov et al. (2025).

In the Galaxy, in addition to the spiral density wave, other factors leading to disk instability are known. We are talking about the gravitational influence of the Large Magellanic Cloud and the passage of other known dwarf satellite galaxies of the Milky Way, such as Sagittarius A, Gaia-Enceladus, and others, through the galactic disk. The manifestation of this influence is observed in the positions of stars as a curvature of the galactic disk---a warp (e.g., Skowron et al. 2019b), and in velocities---warp precession (Poggio et al. 2020). As a result of this influence, periodic perturbations should manifest themselves in stellar velocities of three types---radial, tangential, and vertical.

Based on the data obtained in preparing Fig.~\ref{fig-imr}, we can estimate the absolute value of the difference
$|\Delta\Omega|=|\Omega_p-\Omega_0|$ between the angular velocity of rotation of the spiral pattern $\Omega_p$ and the rotation velocity of the Galaxy $\Omega_0$ at the circumsolar distance $R_0$ using the following relation:
\begin{equation}
|\Omega_p-\Omega_0| = {\Delta \chi_\odot\cdot 1000\over m \Delta t},
\label{MY-chi}
\end{equation}
where $m$ is the number of spiral arms, which is taken to be 4, the phase difference $\Delta \chi_\odot$ is expressed in radians, and the age difference $\Delta t$ is expressed in million years. For all three analyzed samples of OSC, the wavelength $\lambda$ was found to be virtually identical, so calculating the absolute value of $\chi_\odot$ is unnecessary. Thus, to calculate $\Delta \chi_\odot$, we use the found angles $\chi'_\odot$. The absolute value of the difference in angular velocities is included because we make no prior assumptions about the corotation radius, since in the general case we do not know in which direction the wave is moving from the corotation. We estimate the error $\varepsilon_{syst}$ on the right-hand side of expression~(\ref{MY-chi}) as follows:
\begin{equation}
\varepsilon= 1000 \sqrt {
\left({\varepsilon_\chi\over m \Delta t}\right)^2+
\left({\Delta \chi\cdot\varepsilon_t\over m (\Delta t)^2}\right)^2 }
\label{MY-errors}
\end{equation}
for the following accepted characteristic values:
$\varepsilon_\chi=10^\circ \pi/180^\circ,$ and $\varepsilon_t=10$~Myrs.
As a result, we found $|\Delta\Omega|=2.0\pm0.5_{stat}\pm2.3_{syst}$~km/s/kpc as the average of three differences of the form $\Delta \chi'_\odot$.

We calculate the corotation radius based on the ratio obtained by equating the linear rotation velocity of the Galaxy and the found difference in rotation velocities:
\begin{equation}
R_{\rm cor}=R_0\pm|\Delta\Omega|/\Omega'_0 ~.
\label{R-cor}
\end{equation}
Using $\Omega'_0$ from the last column of the table~\ref{t:200}, we find two possible values for the corotation radius
\begin{equation}
\begin{array}{lll}
R1_{\rm cor}= 8.6\pm0.2~\hbox{kpc}\\
R2_{\rm cor}= 7.6\pm0.2~\hbox{kpc},
\label{Rcor-I}
\end{array}
\end{equation}
which indicate that the Sun is very close to corotation.

In the paper by Bobylev et al. (2025), based on an analysis of the amplitudes of the $f_R$ and $f_\theta$ perturbation velocities, found from a sample of masers with VLBI measurements of their trigonometric parallaxes and proper motions, it was shown that there are two possible values of the corotation radius. Namely, $R1_{cor}=9.1\pm0.8$~kpc and $R2_{cor}=6.8\pm0.8$~kpc.
We can see that in the present paper, the range of possible $R_{cor}$ values is significantly narrowed.

\section{CONCLUSION}
We analyzed the kinematics of open star clusters using their spatial and kinematic characteristics from the latest version of the Hunt and Reffert (2024) catalog, in which they were calculated using data from the Gaia DR3 catalog. Average radial velocities are known for a significant number of open clusters in this list. In this paper, we use open clusters for which radial velocity measurement errors do not exceed 8 km/s.

It has been shown that the galactic rotation parameters determined from several samples of OSCs younger than 200 million years are in very good agreement with each other. We note the general solution obtained from 4003 OSCs younger than 200 million years (with an average age of 79 million years) using only their proper motions and distances. Using this approach, the following values for the parameters of the angular velocity of the Galaxy's rotation were found:
$\Omega_0 = 28.99\pm0.11$~km/s/kpc,
$\Omega^{'}_0 = -3.909\pm0.026$~km/s/kpc$^{2}$ and
$\Omega^{''}_0 = 0.5662\pm0.018$~km/s/kpc$^{3}$, where $V_0 = 234.8\pm3.0$~km/s for $R_0 = 8.1\pm0.1$~kpc.

It was established that periodicity in the radial velocities of OSCs is present throughout the analyzed age range of OSCs (younger than 600 million years), and a wave in residual tangential velocities is observed only in the youngest OSCs (younger than 40 million years).

A spectral Fourier analysis of the radial velocities of OSCs was conducted. Using three samples with average ages of 18, 72, and 143 million years, the following values were found:
$\lambda=2.0$~kpc and $f_R=4.3$~km/s,
$\lambda=2.2$~kpc and $f_R=8.2$~km/s,
$\lambda=2.1$~kpc and $f_R=9.6$~km/s, respectively. It was shown that there is a systematic change in the positions of the maxima and minima of the waves in the radial velocities of OSCs depending on the sample age. An analysis of these shifts yielded the absolute value of the difference $|\Delta\Omega|=|\Omega_p-\Omega_0|$ between the angular velocity of the spiral pattern $\Omega_p$ and the rotation velocity of the Galaxy, $|\Delta\Omega|=2.0\pm0.5_{stat}\pm2.3_{syst}$~km/s/kpc. Based on this, two possible values for the corotation radius were estimated: $8.6\pm0.2$~kpc and $7.6\pm0.2$~kpc, indicating that the Sun is very close to the corotation.

\medskip
The authors are grateful to the reviewers for their helpful comments, which contributed to improving the article.

\bigskip\medskip
{REFERENCES}\medskip{\small
\begin{enumerate} \small

 \item
L. H. Amaral,  J. R. D. Lepine, Mon. Not. R. Astron. Soc.   {\bf 286}, 885  (1997).

 \item
T. Antoja, A. Helmi, M. Romero-G\'omez,  et al., Nature  {\bf 561}, 360 (2018).

\item
T. Antoja, P. Ramos, B. Garcia-Conde, et al., Astron. Astrophys. {\bf 673}, A115 (2023)

\item
J.J. Armstrong,  J.C. Tan,  N.J. Wright,  et al.,  Mon. Not. R. Astron. Soc.   {\bf 543}, 2349 (2025).

\item
C. Babusiaux, C. Fabricius, S. Khanna, et al., Astron. Astrophys. {\bf 674}, A32 (2023).

\item
A.T. Bajkova, V.V. Bobylev, Astron. Lett. {\bf 38}, 549 (2012).

\item
A.T. Barnes, K. Morii, J. E. Pineda, et al., arXiv: 2601.20928 (2026).

 \item
V.V. Bobylev, A.T. Bajkova, and S.V. Lebedeva. Astron. Lett.  {\bf 33}, 720 (2007).

 \item
V.V. Bobylev,  A.T. Bajkova, and A.S. Stepanishchev, Astron. Lett. {\bf 34}, 515 (2008).

\item
V.V. Bobylev, A.T. Bajkova, Mon. Not. R. Astron. Soc. {\bf 437}, 1549 (2014). 

 \item
V.V. Bobylev,  A.T. Bajkova, Astron. Lett.  {\bf 42}, 90 (2016).

 \item
V. V. Bobylev,  A. T. Bajkova, Astron. Lett.  {\bf 45},109 (2019). 

\item
V.V. Bobylev, A.T. Bajkova, Astron. Rep. {\bf 65}, 498 (2021). 

\item
V.V. Bobylev, A.T. Bajkova,  Astron. Lett., {\bf 48}, 9 (2022).

\item
V.V. Bobylev, A.T. Bajkova,  Astron. Lett. {\bf 49}, 320 (2023).

\item
V.V. Bobylev, A.T. Bajkova,  A.A. Smirnov, Astron. Lett. {\bf 51}, 278 (2025).

\item
G. A. Gontcharov, A. A. Marchuk,  S. S. Savchenko, et al., Res. Astron. Astrophys.  {\bf 25}, id.125016 (2025).

\item
G. M. Green,  E. Schlafly,  C. Zucker,  et al.,  Astrophys. J. {\bf 887}, 93 (2019).

 \item
W. S. Dias,  J. R. D. Lepine, and B. S. Alessi, Astron. Astrophys. {\bf 376}, с. 441 (2001).

 \item
W. S. Dias,  M. Assafin, V. Florio, et al., Astron. Astrophys. {\bf 446}, 949 (2006).

 \item
W. S. Dias, H. Monteiro, A. Moitinho, et al., Mon. Not. R. Astron. Soc.  {\bf 504}, 356 (2021).

\item
T. C. Junqueira,  C. Chiappini, J. R. D. Lepine, et al., Mon. Not. R. Astron. Soc.  {\bf 449}, 2336 (2015).

\item
M. V. Zabolotskikh,  A. S. Rastorguev, and A. K. Dambis, Astron. Lett. {\bf 28}, 454 (2002).

\item
Y.C.  Joshi,  S. Malhotra, Astron. J. {\bf 166}, 170 (2023).

 \item
D. Camargo, C. Bonatto, and E. Bica,  Mon. Not. R. Astron. Soc.  {\bf 450}, 4150 (2015).

 \item
T. Cantat-Gaudin, A. Jordi, A. Vallenari, et al., Astron. Astrophys. {\bf 618}, A93 (2018).

 \item
T. Cantat-Gaudin, F. Anders, A. Castro-Ginard, et al., Astron. Astrophys.  {\bf 640}, A1  (2020).

\item
Gaia Collab (T. Prusti, et al.), Astron. Astrophys. {\bf 595}, 1 (2016). 

 \item
Gaia Collab ( A.G.A. Brown,  et al.), Astron. Astrophys. {\bf 616}, 1 (2018). 

\item
Gaia Collab (A.G.A. Brown, et al.), Astron. Astrophys. {\bf 649}, 1 (2021). 

 \item
Gaia Collab (A. Vallenari, et al.),  Astron. Astrophys. {\bf 674}, 1 (2023). 

 \item
J. R. D. Lepine,  W. S. Dias, and Y. Mishurov, Mon. Not. R. Astron. Soc.  {\bf 386}, 2081 (2008).

\item
L. Lindegren, S.A. Klioner, J. Hern\'andez,  et al., Astron. Astrophys. {\bf 649}, A2 (2021).

 \item
R.A. Lee, E. Gaidos, J. van Saders, et al., Mon. Not. R. Astron. Soc.  {\bf 528}, 4760 (2024).

 \item
A. V. Loktin,  M. E. Popova, Astron. Rep.  {\bf 51}, 364 (2007).

 \item
A. V. Loktin,  M. E. Popova, Astrophys. Bull.  {\bf 74}, 270 (2019).

 \item
K.L. Luhman, Astron. J. {\bf 165}, 269 (2023).

\item
L Liu, X. Pang, Astrophys. J. Suppl.  {\bf 245}, 32 (2019).

 \item
S. R. Majewski, R. P. Schiavon, P. M. Frinchaboy,  et al., Astron. J. {\bf 154}, 94 (2017).

\item
P. Mr\'oz,  A. Udalski,  D.M. Skowron, et al.,  Astrophys. J. {\bf 870}, L10.  (2019).

\item
M. Perryman, Physics Reports {\bf 1150}, 1 (2026).

 \item
A. E. Piskunov,  N. V. Kharchenko, S. R\"oser,  et al., Astron. Astrophys.  {\bf 445}, 545 (2006).

 \item
E. Poggio, R. Drimmel, R. Andrae, et. al., Nature Astron. {\bf 4}, 590 (2020).

 \item
M. E.  Popova, A. V. Loktin, Astron. Lett.  {\bf 31}, 171 (2005).

\item
S. Ratzenb\"ock,  J.E.  Grosschedl,  T. M\"oller,  et al., Astron. Astrophys. {\bf 677}, A59 (2023).

\item
M.J. Reid, K.M. Menten, A. Brunthaler,  et al., Astrophys. J. {\bf 885}, 131 (2019).

\item
D.M. Skowron, J. Skowron, P. Mr\'oz, et al., Science {\bf 365}, 478 (2019a).

\item
D.M. Skowron, J. Skowron, P. Mr\'oz, et al., Acta Astron. {\bf 69}, 305 (2019b).

 \item
Y. Tarricq, C. Soubiran, L. Casamiquela, et al., Astron. Astrophys.  {\bf 647}, A19  (2021).

 \item
T. Tepper-Garcia, J. Bland-Hawthorn,  T. R. Bedding,  et al., Mon. Not. R. Astron. Soc.  {\bf 542}, 1987 (2025).

\item
Q. Feng, Y. Huang, H. Zhang,  and J. Liu,   Mon. Not. R. Astron. Soc.  {\bf 546}, id.stag011 (2025).

\item
P.N. Fedorov  A. M. Dmytrenko, V. S. Akhmetov, et al., arXiv: 2511.22295 (2025).

 \item
C.J. Hao, Y. Xu, L.G. Hou,  et al., Astron. Astrophys. {\bf 652}, 102 (2021).

\item
C. Hamilton, A. Mummery, and J. Bland-Hawthorn, arXiv: 2602.06182 (2026).

\item
E.L. Hunt, S. Reffert, Astron. Astrophys. {\bf 646}, A104 (2021).

\item
E.L. Hunt, S. Reffert, Astron. Astrophys. {\bf 673}, A114 (2023).

\item
E.L. Hunt, S. Reffert, Astron. Astrophys. {\bf 696}, A42 (2024).

 \item
N.V. Kharchenko,  A. E. Piskunov, S. R\"oser,  et al., Astron. Astrophys. {\bf 438}, 1163 (2005).

 \item
N.V. Kharchenko,  R.-D. Scholz, A. E. Piskunov,  et al., Astron. Nachr. {\bf 328}, 889 (2007).

 \item
N.V. Kharchenko,  A. E. Piskunov, E. Schilbach,  et al., Astron. Astrophys. {\bf 558}, A53 (2013).

 \item
D. Horta, A. M. Price-Whelan, S.E. Koposov, et al., Xiv: 2601.18876 (2026).

\item
R. Chiba, N. Frankel, and C. Hamilton, Mon. Not. R. Astron. Soc. {\bf 543}, 2159 (2025).

 \item
R.-D. Scholz, N. V. Kharchenko, A. E. Piskunov,  et al., Astron. Astrophys. {\bf 581}, A39 (2015).

 \item
R. Sch\"onrich, J.J. Binney, and W. Dehnen, Mon. Not. R. Astron. Soc. {\bf 403}, 1829 (2010).

 \item
P. Yamsiri, J. Bland-Hawthorn, and T. Tepper-Garcia, arXiv: 2602.05296 (2026).

 \end{enumerate} }
 \end{document}